\documentclass[oneside,twocolumn,english]{article}

\usepackage{geometry}
\usepackage{babel}
\newcommand{\noun}[1]{\textsc{#1}}
\begin{document}

\twocolumn[\hsize\textwidth\columnwidth\hsize\csname
@twocolumnfalse\endcsname

\title{\Large \bf {Relational EPR}}

\author{Matteo Smerlak$^{\dagger}$, Carlo Rovelli$^{\ddagger}$
\\ \\  {\it $^{\dagger}$\'Ecole Normale Sup\'erieure de Lyon, F-69364 Lyon, EU} 
\\{\it $^{\ddagger}$Centre de Physique Th\'eorique de Luminy, F-13288 Marseille, EU}
\\ \\Ê
}
\vspace{3\baselineskip}

\maketitle

{\noindent We study the EPR-type correlations from the perspective of the relational interpretation 
of quantum mechanics. We argue that these correlations do not entail any form 
of  ``non-locality", when viewed in the context of this interpretation. 
The abandonment of strict Einstein 
realism implied by the relational stance permits to reconcile quantum 
mechanics, completeness, (operationally defined) separability,  and locality.}

\vspace{\baselineskip}

DOI: 10.1007/s10701-007-9105-0
 \vspace{1cm}

 ]

\section{Introduction}

EPR-type experiments, championed by Aspect   
\emph{et al.\,} \cite{key-20,key-28},
are often interpreted as empirical evidence for the existence of 
a somewhat mysterious ``quantum non-locality''.  For instance, 
Chris Isham concludes his beautiful exposition 
of the EPR debate with the words ``[...] we are obliged either to stick 
to a pragmatic approach or strict instrumentalist interpretation, or else to accept the 
existence of a strange non-locality that seems hard to reconcile with
our normal concepts of spatial separation between independent entities"  \cite{key-19}. 
In spite of seven decades of reflection on this problem, 
leading to considerable sharpening in its characterization 
 \cite{altri,altri2},
the precise nature of this non-locality ---which does not appear to be
usable to transmit information, 
nor does make quantum theory incompatible with special 
relativity--- remains rather elusive. 

In recent years, a novel point of view on quantum theory, denoted 
Relational Quantum Mechanics (RQM), has been discussed by
some authors \cite{key-10,key-43,key-48,key-44,key-40,key-41,key-42}. 
In this paper, we argue that in the context 
of this interpretation, it is not necessary to abandon 
locality in order to account for EPR correlations. 
From the relational perspective, the apparent ``quantum 
non-locality'' is a mistaken illusion caused by the error of 
disregarding the quantum nature of \emph{all} physical systems. 

The price for saving locality is the weakening of
realism which is at the core of RQM. This ontological move, as radical as it may appear at first sight, is actually implicit in the historical evolution of the EPR debate. 

In the original 1935 
article \cite{key-25}, the EPR argument was conceived as an 
attack against the description of measurements in Copenhagen 
quantum theory, and a criticism of the idea that Copenhagen
QM could be a \emph{complete} description of reality. 
Locality and a strong form of realism were given for granted
by EPR, and completeness was argued to be incompatible
with quantum-mechanical predictions. 

With Bell's contribution  
\cite{key-22}, which showed that EPR correlations are incompatible with the existence 
of a hypothetical \emph{complete} local 
realist theory, the argument has been mostly reinterpreted as a direct challenge to 
``local realism".  Proofs of non-locality  have been  then 
developed by a number of other authors, using increasingly  
weaker assumptions (\cite{altri,altri2}, and references therein), in particular,
dropping the need of assuming the existence of a hidden--variable theory.

On the other hand, the Kochen-Specker theorem   \cite{key-23} 
has questioned the very possibility of uncritically ascribing 
 ``properties'' to a quantum system. From this perspective, the problem  of locality 
moves to the background, replaced by a mounting critique 
of strongly objective notions of reality (see for instance \cite{key-29}). Here we take 
this conceptual evolution to what appears to us to be 
its necessary arriving point:  the possibility of reading EPR-type 
experiments as a challenge to Einstein's strong realism, rather 
than locality.

To be sure, the philosophical implications of RQM, especially for what concerns 
realism, are heavy\footnote{We refer the reader puzzled by these philosophical 
implications to \cite{key-10}, where 
the position of RQM in the landscape of current views on 
quantum theory is discussed in detail. See also the discussion in 
Section 5.6 of \cite{libro}.}.  We shall briefly comment on 
these in Sec 4.4.  However, the purpose of this paper is not to 
defend explicitly the relational interpretation of quantum theory, but only to remark that, if one adopts this view, the
disturbing non-local features of EPR-like correlations
disappear.

Similar criticisms to the notion of ``quantum non-locality" have been recently expressed by a number of authors  
\cite{key-46,key-35,key-37,key-38}. In particular, in a recent article \cite{key-34},
Asher Peres concludes his analysis of the EPR problem with a general 
statement, which, as we shall see below, is precisely the ground assumption 
of RQM. Thus, if we are inclined to accept RQM as a way to make sense
of quantum theory, the EPR correlations can be interpreted as 
supporting this point of view.

\section{Relational quantum mechanics, locality and separability}

The relational approach claims that a number of confusing puzzles 
raised by Quantum Mechanics (QM) result from the unjustified use 
of the notion of objective, absolute, `state' of a physical system,
or from the notion of absolute, real, `event'. 

The way out from the confusion suggested by RQM consists in 
acknowledging that \emph{different
observers can give different accounts of the actuality of the same physical property} \cite{key-10}.  This fact implies that the occurrence of an event is not 
something absolutely real or not, but it is only real 
in relation to a specific observer. Notice that, in this context, an observer can be \emph{any} physical system.

Thus, the central idea of RQM is to apply Bohr and Heisenberg's key intuition 
that ``no phenomenon is a phenomenon until it is an observed phenomenon'' 
to each observer independently. This description of physical reality, though fundamentally
fragmented, is assumed in RQM to be the best possible one, {\em i.e.} 
to be complete \cite{key-10}: 
\begin{quote}\em ``Quantum mechanics is a theory about the physical 
description of physical systems relative to other systems, and this is a 
complete description of the world".\end{quote}

\subsection{RQM and Physical Reality}\label{realism}

In the context of the EPR debate, {\em realism} is taken as 
the assumption that, in Einstein's words  \cite{key-33}, 
\begin{quote}\em 
``there exists 
a physical reality independent of substantiation and 
perception".
\end{quote}  
We call this assumption ``Einstein's realism".\footnote{This 
simplistic definition does not do justice
to Einstein's subtle position concerning physical reality, which 
is far more instrumental and programmatic than metaphysical:
see \cite{key-31} for an insightful analysis. We use this 
definition here it for the sake of the discussion.}
 RQM  departs from such strict realism.  In RQM, 
physical reality is taken to be
formed by the individual \emph{quantum events} (facts%
\footnote{ ``1.1 The world is the totality of facts, not of things"
\cite{key-39}.}%
) through which interacting systems (objects%
\footnote{ ``2.01 An atomic fact is a combination of objects (entities, things).
2.011 It is essential to a thing that it can be a constituent part of an atomic fact"
\cite{key-39}.}%
) affect one another. 
Quantum events are therefore assumed to exist only in interactions%
\footnote{ ``2.0121 There is no object that we can imagine 
excluded from the possibility of combining with others" \cite{key-39}.}
and (this is the central point) {\em the character of each quantum event 
is only relative to the system involved in the  interaction}. In particular,
which properties any given system $S$ has is only relative
to a physical system $A$ that interacts with $S$ and is affected by these
properties. 

If $A$ can keep track
of the sequence of her past interactions with $S$, then $A$  
has \emph{information} about $S$, in the sense that $S$ and $A$'s degrees of freedom are \emph{correlated}. 
 According to RQM, this relational information exhausts the content 
of any observer's description of the physical world.   

Michel Bitbol
proposes to qualify this approach as a \emph{meta-description} \cite{key-40}:
RQM is the set of rules specifying the form of any such physical 
description. In that sense, RQM is faithful to Bohr's epistemological 
position, as presented for instance in \cite{key-12}:
\begin{quote}\em
``It is wrong to think that the task of physics is to find out how nature
is. Physics concerns what we can say about nature."
\end{quote}
Still, RQM adds an essential twist to this position of Bohr. 
For Bohr, the ``we" that can say something about nature
is a preferred macroscopic classical apparatus that escapes
the laws of quantum theory: facts, namely results of quantum measurements,
are produced interacting with this classical observer. 
In RQM, the preferred observer is abandoned. Indeed, it is a fundamental assumption of this approach that nothing distinguishes
\emph{a priori} systems and observers: any 
physical system provides a potential observer.
Physics concerns what can be said about nature on the basis of the 
information that \emph{any} physical system can, in principle, have.  
The preferred Copenhagen observer is relativized into the multiplicity of
observers, formed by \emph{all} possible physical systems, and therefore it 
no longer escapes the laws of quantum mechanics\footnote{An observer, in the sense used here, does not need to be, say ``complex", or
even less so ``conscious".   An atom interacting with another atom 
can be considered an observer.  Obviously this does not mean that one
atom must be capable of storing the information about the other atom, and
consciously computing the outcome of its future interaction with it;
the point is simply that the history of its past interaction \emph{is} is
principle sufficient information for this computation. 
}. 

Different observers can of course exchange information. However, this
exchange is itself a quantum mechanical interaction. 
An exchange of information is a quantum measurement 
performed by one observing system $A$  upon another observing
system $B$. As we shall see, it is the disregard of this fact of nature
that creates the illusion of the EPR non-locality.

\subsection{The Physical Meaning of $\psi$}

The existence in regularities in natural phenomena, that is, 
laws of nature, means that we can predict future events on 
the basis of past events.  More precisely, the outcome of future 
interactions of an ``observing" system $A$ with an observed
system $S$ can be predicted on the basis of the information 
acquired via past interactions.   The tool for doing this is the 
quantum state $\psi$ of $S$.

The state $\psi$ that we associate with a system $S$ is therefore, 
first of all, just a coding of the outcome of these previous interactions 
with $S$.  Since these are actual only with respect to $A$, the state 
$\psi$ is only relative to $A$: \emph{$\psi$ 
is the coding of the information that $A$ has about $S$}.  Because of this 
irreducible epistemic character, $\psi$ is 
but a relative state, which cannot be taken to be an objective property 
of the single system $S$, independent from $A$.  Every state of quantum 
theory is a relative state.%
\footnote{From this perspective, probability needs clearly to be interpreted
subjectively.  On a Bayesian approach to QM, see 
\cite{key-35,key-36};
for a more general defense of Bayesian probabilities in science, and 
a discussion of the relevance of this point for the EPR debate, 
see \cite{key-37}.} 

On the other hand,
the state $\psi$ is a tool that can be used by $A$ to \emph{predict} future outcomes
of interactions between $S$ and $A$. In general 
these predictions depend on the time $t$ at which the interaction will 
take place. In the Schr\"odinger picture this
time dependence is coded into a time evolution of the state $\psi$ itself. In this picture, 
there are therefore two distinct manners in which $\psi$ can evolve: (i) in a 
discrete way, when $S$ and $A$ interact, in order for the information
to be adjusted, and (ii) in a continuous way, to reflect the time
dependence of the probabilistic relation between past and future events. 

From the relational perspective the Heisenberg picture appears far more natural: $\psi$ codes the information that can be extracted from past interactions and has no explicit dependence on time; it is adjusted 
only as a result of an interaction, namely as a result of a new quantum event relative to the observer. If physical reality is the set of these bipartite interactions, and nothing else, our description of dynamics by means of relative states should better mirror this fact: discrete changes of the relative state, when information is updated, and nothing else. What evolves with time are the operators, whose expectation values code the time-dependent probabilities that can be computed on the basis of the past quantum events.%
\footnote{This was also Dirac's opinion: in the first edition 
of his celebrated book on quantum mechanics, Dirac uses 
Heisenberg states (he calls them relativistic) \cite{key-47}. 
In later editions, he switches to Schr\"odinger states, 
explaining in the preface that it is easier to calculate 
with these, but it is ``a pity" to give up Heisenberg 
states, which are more fundamental. In what was
perhaps his last public seminar, in Sicily, Dirac used 
a single transparency, with just one sentence: ``The 
Heisenberg picture is the right one".}

To summarize, two distinct aspects of physical information, \emph{epistemic} 
and \emph{predictive}, are subsumed under the notion 
of (relative) quantum state;  amending Bohr's epistemology, we can say that QM is the 
theory of logical relations between the two.  

\subsection{Locality}

We call \emph{locality} the principle demanding that 
\emph{two spatially separated events cannot have instantaneous 
mutual influence}.   We will argue that
this is not contradicted by EPR-type correlations, if we take the relational
perspective on quantum mechanics.

Locality is at the very roots of RQM, in the observation
that different observers (in general distant from one another) can have different descriptions of the same system. 

As emphasized by Einstein, it is locality that makes possible the individuation
of physical systems, including those we call observers%
\footnote{``Without the assumption of the mutually independent existence
(the `being-thus') of spatially distant things, an assumption which
originates in everyday thought, physical thought in the sense familiar
to us would not be possible. Nor one does can see how physical laws
could be formulated and tested without such a clean separation.''
Quoted in \cite{key-24}, where this point is discussed in depth.
}. From the RQM perspective, this observation amounts to acknowledging the relative character of actuality. Indeed, recall that a property of $S$ is actual relative to $A$ only if substantialized in a correlation between $A$ and $S$. This (epistemic) correlation is always constrained by the speed of light, so that distant observers are bound to have different information on a given system: they do not describe reality univocally. 

An indication of this fact is in the well-known difficulty 
of describing and interpreting the relativistic transformation 
law of the wave function, when measurements involve observers 
in relative motion \cite{key-32}.

Even beyond its foundational role in relativistic field
theories, locality constitutes, therefore, the base of the
relational methodology: an observer cannot, and must not, account
for events involving systems located out of its causal neighborhood
(or light-cone).%
\footnote{We can take this observation as
an echo in fundamental physics of the celebrated: 
``7. Whereof one cannot speak, thereof one must be silent" 
 \cite{key-39}.}
 
These remarks lead us to the following reformulation of the locality principle, in which the relational perspective is made explicit: \emph{relative to a given observer, two spatially separated events cannot have instantaneous mutual influence}.

The idea that locality imposes
a relativization in the description of reality
is certainly not new: it is precisely the
physical content of special relativity. When we say that simultaneity
is relative, we mean that distant observers \emph{cannot} 
take note of a given event instantaneously, and thus ascribe 
it the same time as their own.   The meaning of the adjective 
``relative'' in the RQM notion of  ``relative state'' is therefore very 
similar to the meaning of ``relative" in special relativity. It
is the translation of the impossibility of principle to transmit 
information faster than light ---and without a physical interaction.
To stress the analogy, we can say that 
the conceptual difficulties raised by the interpretation
of the Lorentz transformations before 1905 came from the lack of appreciation
of  the epistemic nature of simultaneity.

\subsection{Separability}

Another concept playing an important role in the EPR 
discussions is \emph{separability}. An option that saves 
a (weakened) form of locality  is, according to some, 
to assume that entangled quantum objects are ``not-separable". 
Aspect, for instance, teaches that ``a pair of twin entangled photons 
must in fact be regarded as a single, \emph{inseparable} system, 
described by a global quantum state'' \cite{key-9}.  If this is just
a restatement  of the existence of correlations, and the consequent 
impossibility of assigning well-defined independent states to the 
photons, this is unquestionable. But if this is meant to provide an ontologically satisfactory explanation of the mysterious EPR correlations, then it clearly misses its point, since experiments do perform measurements on distinct photons. In fact, this rather strange notion, where two physical entities
are actually a single system, indicates, in our opinion, the difficulty 
to reconcile realism, locality and quantum theory. 
We argue below that the abandonment of Einstein's strict realism 
allows one to exempt himself from this type of intellectual acrobatics.

Let us instead choose the following definition of separability: 
two physical systems $S_{1}$ and $S_{2}$ are \emph{separable}
if there exists a complete set of observables (in the sense of Dirac) of 
the compound system $S_{1}+S_{2}$ whose values can be actualized 
by measurements on $S_{1}$ or $S_{2}$ only. 
Such observables are called \emph{individual} \emph{observables};
the others are called \emph{collective} \emph{observables}.

This notion of separability is equivalent to a minimal operational
definition of subsystems of a composite system. It is deliberately
weak (and in the end trivial); any stronger definition testifies
to some unnecessary unease.

\section{The EPR argument}

Let us start from Einstein's formulation%
\footnote{For a detailed discussion of Einstein's position on the question of locality,
appreciably different from the one presented in the original EPR article, see
\cite{key-24,key-31} and below.} 
 of the EPR argument, and then analyze its later evolutions. 

\subsection{Reminder of the Experiment (in Bohm's setting)}

Consider a radioactive decay, producing two spin-half particles, and
call them $\alpha$ and $\beta$. Suppose that some previous measurement
ensures that the square of the total spin of the two particles equals zero ---which corresponds, in the spectroscopic vocabulary, to the
singlet state. The particles $\alpha$ and $\beta$ leave the source
in two different directions, reaching two distant detectors $A$ and
$B$, which measure their spin in given directions.

\subsection{Einstein's Version of the EPR Argument}

According to standard QM, the measurement of an observable provokes the projection
of the system's state onto the eigen\-space associated with the obtained
eigenvalue. In the case of the singlet, the state can be equivalently
decomposed on the eigenbasis of the spin in two different directions,
say $z$ and $x$:
\begin{eqnarray}
\left|\psi_{\rm singlet}\right\rangle &=&\frac{1}{\sqrt{2}}\left(\left|\downarrow\right\rangle _{\alpha}\left|\uparrow\right\rangle _{\mathbf{\beta}}-\left|\uparrow\right\rangle _{\alpha}\left|\downarrow\right\rangle _{\mathbf{\beta}}\right)\nonumber \\
&=&\frac{1}{\sqrt{2}}\left(\left|\rightarrow\right\rangle _{\alpha}\left|\leftarrow\right\rangle _{\mathbf{\beta}}-\left|\leftarrow\right\rangle _{\alpha}\left|\rightarrow\right\rangle _{\mathbf{\beta}}\right).
\end{eqnarray}
Depending on whether the observer $A$ measures the spin of
$\alpha$ in the direction $z$ or $x$, the second particle $\beta$
finds itself in an eigenstate of $S^{z}$ or $S^{x}$. In either case,
the property of having a definite spin in one direction is uniquely
determined for $\beta$, hence is real, since, according to Einstein's 
realism,
\begin{quote}\em 
If, without in any way disturbing a system, we can predict with certainty the value
of a physical quantity, then there exists an element of physical reality corresponding to
this physical quantity. \cite{key-25}
\end{quote}
But according to
the principle of locality, the choice made by $A$ cannot have an influence
on $\beta$, which is space-like separated from $A$. 
Therefore, in order to accommodate both possibilities
it is necessary for the spin in  \emph{both} directions to be uniquely 
determined.  But this is more physical information than the one contained
in a vector in the Hilbert space of the states of $\beta$. 
Hence there exist real properties not described by quantum mechanics. 
Completeness of quantum mechanics, namely one-to-one correspondence 
between the mathematical objects used to describe the state of a system and
its real state, is disproved.

\subsection{The Question of Locality} 

Einstein and his collaborators in the EPR paper had no reasons whatsoever to question
locality, or realism. The first was one of the pillars of Einstein's 
major achievements. The second was a philosophical assumption to which
science was obviously immensely indebted. 

But Bell's work showed that the simplest interpretation of EPR correlations
as an indication that quantum mechanics is incomplete was not tenable: 
any hypothetical complete classical dynamics yielding the same correlations 
as quantum mechanics violates locality. If the quantum predictions are correct, then
a realistic local theory seems impossible.

Of course, the possibility was still open that QM was simply not yielding the 
correct physical predictions. But this last possibility has been ruled out by the 
experimental work of Aspect et al, leaving, it seems,
only two possible interpretations of the EPR correlations: either as 
a manifestation of non-locality, as commonly assumed, or, 
as a challenge to strong realism.  It is this second possibility, we argue
here, which is made concrete by RQM.

\subsection{Relational Critique} 

Einstein's argument relies on the strongly realistic hypothesis that 
\emph{the actual properties of the particles (the ``real state of affairs'') 
revealed by the detectors are observer-independent.} It is this hypothesis 
that justifies the ascription of a definite, objective, state to each particle, at 
every instant of the 
experiment: in Einstein's account, when $B$ measures the spin of $\beta$, 
the measured value {\em instantaneously acquires  an objective existence 
also relative to $A$}. 

This hypothesis, namely that  when $B$ measures the spin of $\beta$, 
the measured value instantaneously acquires  an objective existence 
that can be considered absolute, is common to all the analyses that 
lead to an interpretation of the EPR correlations as a manifestation
of non-locality \cite{altri,altri2}. 

But this hypothesis is not operationally justified: nothing enables $A$ to 
know the outcome of the measure 
carried out by $B$ on $\beta$, unless $A$ measures the state of $B$.
$A$ cannot measure the state of $B$ instantaneously, 
precisely because of locality: $B$ is far away.

From the relational perspective, what is missing in 
Einstein's quotation above, as well as in all later analyses of the
EPR correlations, is the distinction between
``elements of physical reality" (quantum events) relative to $A$ and ``elements 
of physical reality" relative to $B$.

Observer $A$ can of course measure the
state of $B$ (or, for that matter, $\beta$), but only when $A$ is back
into causal contact with $B$ \cite{key-50}. This is, needless to say, in the future light-cone
of $A$, and therefore poses no challenge for locality.  In other words,
Einstein's reasoning requires the existence of a hypothetical
super-observer that can instantaneously measure the state of $A$
and $B$. It is the hypothetical existence of such nonlocal super-being, and 
not QM, that violates locality.

Let us look at the origin of the illusion of
non-locality more in detail. Suppose that
$A$ measures a spin component of $\alpha$ at time $t_0$, and $B$ 
measures a spin component of $\beta$ at time $t_0'$. Einstein's 
ingenious counterfactual argument works under the assumption 
that locality prevents any causal 
influence of $A$'s measurement on $B$'s ($A$'s choice 
of measuring the spin along $z$ or along $x$ cannot
affect the $B$ measurement, hence we can counterfactually
join the consequences of the two alternatives). But for such 
counterfactuality to be effective, there has to exist an objective 
``element of reality" which is unaffected by $A$'s actions.  If one 
acknowledges that $B$'s state of affairs is \emph{a priori} 
undefined for $A$, then bringing $B$ into the argument is 
useless, because then what would be actualized by $A$'s 
measurement of the spin of $\alpha$ along one direction 
would be \emph{relative to $A$} only. 
In fact, Einstein implicitly assumes that $B$ is a \emph{classical} 
system, recording \emph{objective} values in its ``pointer 
variables".  That is, even if $A$ can't see the position of $B$'s pointer 
variables before a later time $t_1$, this position has nevertheless 
a \emph{determined} position since $t_0'$\footnote{A similar 
implicit hypothesis, the ``retrodiction principle", was pointed out by 
Bitbol in \cite{key-46}.}. Thus, the properties of $\beta$ become 
actual when it interacts with $B$ at time $t_0'$, indeed 
substantiating the non-local EPR correlations between 
distant locations. Thus, {\em it is the assumption that $B$ 
is classical and fails to obey quantum theory that creates 
EPR non-locality.}
 
But all systems are quantum: there are no intrinsically classical 
systems. Hence the hypothesis that $B$ does not obey quantum
theory is physically incorrect.  It is this mistaken hypothesis 
that causes the apparent violation of locality. 

In other words, in the sequence of events which is real for $A$
there is no definite quantum event regarding $\beta$ at 
time $t_0$, and therefore no element of reality generated 
non-locally  at time $t_0$ in the location where $B$ is. Hence 
Einstein's argument cannot even begin to be formulated.

What changes instantaneously at time $t_0$, for $A$, is 
not the objective state of $\beta$, but only its (subjective) 
relative state, that codes the information that $A$ has 
about $\beta$. This change is unproblematic, for the same 
reason for which my information about China changes 
discontinuously any time I read an article about China in
the  newspaper. 
Relative to $A$, $\beta$ is not affected by this change 
because there is no $\beta$-event happening at time 
$t_0$. The meaning of the sudden change in the state
of $\beta$ is that, as a consequence of her measurement
on $\alpha$, $A$ can predict outcomes of  \emph{future}
measurement that $A$ herself might do on $\beta$, or
on $B$.   

Of course the price to pay for this solution of the puzzle is that 
the sequence of events as described by $B$ is different from
what it is as described by $A$. For $B$, there \emph{is} a 
quantum event of $\beta$ at time $t_0'$ and there is 
\emph{no} quantum event regarding $\alpha$ at 
time $t_0'$. But the core assumption of RQM is that 
quantum events relative to distinct observers cannot be
simply juxtaposed.

Finally, let us add one remark on the later arguments supporting
the idea of non-locality in QM \cite{altri,altri2}.  Some
of these works are based on a weaker form of realism than 
the one of Einstein or Bell. However, they all still maintain the assumption 
that there is \emph{an objective element of reality in the simultaneous
realization of the measurements  of $\alpha$ and $\beta$
at space-like separated locations}.  For instance, in its second premise,
Stapp demands that ``experimental outcomes
that have already occurred in an earlier region [...] can be considered to
be fixed and settled independently from which experiment will be chosen and performed
later in a region space-like separated from the first."  \cite{altri2} This is
precisely the assumption 
questioned in RQM. 

\section{Relational discussion of the EPR experiment}

We shall now present a relational discussion of the EPR experiment,
compatible with locality. But first,
let us get rid of the problem of separability: in the EPR experiment,
the two entangled systems interact with two different observers. 
Incontestably, both get definite outcomes during these complete 
measurements (in the sense of Dirac). Hence, the particles
are separable. Fine - one might say - but what about the EPR correlations?

\subsection{Individual Measurements}

Say that $A$ measures the spin of $\alpha$ in the direction $n$
at time $t_0$. This is an  individual observable, denoted $S_{A\alpha}^{n}$.
Suppose $B$ measures the spin of
$\beta$ in the direction $n'$ at time $t_0^{'}$ (individual observable 
$S_{B\beta}^{n'}$).  Let us denote $\epsilon_{A\alpha}$ and $\epsilon_{B\beta}$ ($\epsilon=\pm1$)
the corresponding outcomes. Because $A$ and $B$ are space-like separated,
there cannot exist an observer with respect to which both of these
outcomes are actual, and therefore it is meaningless to compare 
$\epsilon_{A\alpha}$ and $\epsilon_{B\beta}$: $A$'s outcome is 
fully independent from $B$'s, and \emph{vice versa.}

\subsection{EPR Correlations}

But these individual measures do not exhaust all possibilities. In
the EPR experiment, the composite system $\alpha+\beta$ is assumed
to be in the singlet state. From the relational point of view, this
means that some observer, say $A$ herself, has the information
that the total spin of $\alpha+\beta$ equals zero. That is,
it has interacted with the composite system in the past and
has measured the square of the total spin. Let us call this 
\emph{collective} observable $S_{A,\alpha+\beta}^{2}$.

The measurement of $S_{A\alpha}^{n}$ brings
new information to $A$. It determines the change of the relative state
of $\alpha$. Notice that $A$'s knowledge about $\alpha$ changes 
(epistemic aspect), and, at the same time, $A$' predictions 
concerning future change (predictive aspect). 
For instance, $A$ becomes able to predict with certainty the value 
of $S_{A\alpha}^{n}$ if the interaction is repeated.

But there is another observable whose value QM enables $A$ 
to predict: $S_{A\beta}^{n'}$, namely the measurement that $A$ can 
perform on $\beta$ at the time $t_1$, when $\beta$ is back into
causal contact with $A$.   For instance, if  
\begin{equation}
S_{A,\alpha+\beta}^{2}=0\hspace{1em} {\rm and}\hspace{1em} Ê S_{A\alpha}^{n}=\epsilon,
\end{equation}
then QM predicts
\begin{equation}
S_{A\beta}^{n}=-\epsilon. 
\end{equation}
That is, the knowledge
of the value of the collective observable $S_{\alpha+\beta}^{2}$
plus the knowledge of the individual observable $S_{A\alpha}^{n}$ 
permit to 
predict the future outcome of the individual observable 
$S_{A\beta}^{n}$: it is 
\emph{this} type of inference which constitutes the 
``EPR correlations''. It concerns a sequence of causally connected 
interactions.

\subsection{Consistency}

Let us bring $B$ back into the picture. It is far from the spirit of RQM to 
assume that each observer has a ``solipsistic" picture of reality,
disconnected from the picture of all the other observers. In fact, the very
reason we can do science is because of the consistency we find
in nature: if I see an elephant and I ask you what you see, I expect
you to tell me that you too see an elephant.  If not, something 
is wrong.  

But, as claimed above, any such conversation about 
elephants is ultimately an interaction between quantum systems.
This fact may be irrelevant in everyday life, but disregarding it 
may give rise to subtle confusions, such as the one leading
to the conclusion of non-local EPR influences. 

In the EPR situation, $A$ and $B$ can be considered two
distinct observers, both making measurements on $\alpha$
and $\beta$.  The comparison of the results of their 
measurements, we have argued, cannot be instantaneous, that
is, it requires $A$ and $B$ to be in causal contact. More
importantly, with respect to $A$, $B$ is to be considered as 
a normal quantum system (and, of course, with respect to $B$, 
$A$ is a normal quantum system). So, what happens if $A$ 
and $B$ compare notes?  Have they seen the same elephant? 

It is one of the most remarkable features of quantum mechanics 
that indeed it automatically guarantees precisely the 
kind of consistency that we see in nature \cite{key-10}. Let us illustrate
this assuming that both $A$ and $B$ measure the spin in
the same direction, say $z$, that is $n=n'=z$. 

Since $B$ is a quantum system, there will be an observable 
$S^n_{AB}$ corresponding to $B$'s answer (at time $t_1)$ to the question
``which value of the spin have you measured?".  That is,  
$S^n_{AB}$ is the observable describing the pointer variable
in the detector $B$.  Then consistency demands that: 

(i) If ${A}$ measures $S^n_{AB}$ after having measured 
$S_{A\beta}^{n}$, 
she will get 
\begin{equation}
S^n_{AB} = S_{A\beta}^{n}. 
\end{equation}

(ii) If a third observer $C$, who has the prior information that measurements have been performed by $A$ and $B$, measures at a later time the two pointer variables:  $S^n_{CA}$ and $S^n_{CB}$ then
\begin{equation}
S^n_{CB} = - S_{CA}^{n}. 
\end{equation}

But this follows from standard QM formalism, because
an interaction between $\beta$ and $B$ that can be interpreted
as a measurement is an interaction such that the state (1) and the
initial state of $\alpha$, $\beta$ and $B$ evolve into the state (relative to $A$)
\begin{equation}
\left|\psi\right\rangle_{\alpha+\beta+B}^{(A)}=\frac{1}{\sqrt{2}}\left(\left|\downarrow\right\rangle _{\alpha}\left|\uparrow\right\rangle _{\mathbf{\beta}}
\left|\uparrow\right\rangle _{{B}}
-\left|\uparrow\right\rangle _{\alpha}
\left|\downarrow\right\rangle _{\mathbf{\beta}}
\left|\downarrow\right\rangle _{{B}}
\right)
\end{equation}
with obvious notation. Tracing out the state of $\alpha$ that plays no role here, we get the density matrix 
\begin{equation}
\rho_{\beta+B}^{(A)}=\frac{1}{2}\left(\left|\uparrow\right\rangle _{\beta}\left|\uparrow\right\rangle _{B}\left\langle \uparrow\right|_{\beta}\left\langle \uparrow\right|_{B}+\left|\downarrow\right\rangle _{\beta}\left|\downarrow\right\rangle _{B}\left\langle \downarrow\right|_{\beta}\left\langle \downarrow\right|_{B} \right).
\end{equation}
from which (4) follows immediately.  Similarly, the state of the ensemble of the four systems 
$\alpha,\beta,A,B$, relative to $C$ evolves, after the two interactions at time $t_0$ into the state 
\begin{equation}
\left|\psi\right\rangle\!_{\alpha+\beta+A+B}^{(C)} =\!\frac{1}{ \sqrt{2}}\!\left(\left|\downarrow\right\rangle _{\alpha}\left|\uparrow\right\rangle _{\mathbf{\beta}}
\left|\downarrow\right\rangle _{{A}}
\left|\uparrow\right\rangle _{{B}}\! -\! \left|\uparrow\right\rangle _{\alpha}
\left|\downarrow\right\rangle _{\mathbf{\beta}}
\left|\uparrow\right\rangle _{{A}}
\left|\downarrow\right\rangle _{{B}}
\!\right)
\end{equation}
again with obvious notation. Tracing out the state of $\alpha$ and $\beta$, we get the density matrix 
\begin{equation}
\rho_{A+B}^{(C)}=\frac{1}{2}\Big(\left|\downarrow\right\rangle _{A}\left|\uparrow\right\rangle _{B}\left\langle \downarrow\right|_{A}\left\langle \uparrow\right|_{B}+\left|\downarrow\right\rangle _{A}\left|\uparrow\right\rangle _{B}\left\langle \downarrow\right|_{A}\left\langle \uparrow\right|_{B}\Big).
\end{equation}
which gives  (5) immediately. (For a similar argument, see \cite{vecchi}.) 
It is clear that everybody sees the same elephant. 
More precisely: everybody hears everybody else stating that they see the
same elephant they see.  This, after all, is a sound definition of objectivity. 

\subsection{An Objection}

An instinctive objection to the RQM account of the above situation
is the following.  Suppose that at a certain time the following happens
\begin{quotation} ($\star$) \ \em $A$ observes the spin
in a given direction to be  $\uparrow$ and  $B$ observes the spin
in the same direction to be also $\uparrow$.  
\end{quotation} 
Agreement with quantum theory demands that when later interacting with $B$, $A$ will necessarily finds $B$'s pointer variable indicating that the measured spin was $\downarrow\;$.  This implies that what $A$ measures about $B$'s information ($\downarrow$) is unrelated to what $B$ has actually measured ($\uparrow$). The conclusion appears to be that each observer sees a completely different world, unrelated to what any other observer sees: $A$ sees an elephant and hears $B$ telling her about an elephant, even if $B$ has seen a zebra. Can this happen in the conceptual framework of RQM?  
 
The answer is no. The reason is subtle and lies at the core of RQM.    

The founding postulate of RQM stipulates that we shall not deal with properties of systems in the abstract, but only of properties of systems relative to \emph{one} system.  In particular, we can never juxtapose properties relative to different systems.  If we do so, we make the same mistake as when we simultaneously ascribe position and momentum to a particle. In other words, RQM is \emph{not} the claim that reality is described by the \emph{collection} of all properties relative to all systems. This collection is assumed not to make sense.  Rather, reality admits one description per (observing) system, each being internally consistent.

In turn, any given system can be observed  by another system.  RQM is, in a sense, the stipulation that we shall not talk about anything else than that, and the observation that this scheme is sufficient for describing nature and our own possibility of exchanging information about nature (hence circumventing solipsism).   

So, the case ($\star$) can never happen, because it does not happen either with respect to $A$ or with respect to $B$.  The two sequences of events (the one with respect to $A$ and the one with respect to $B$) are distinct accounts of the same reality that cannot and should not be juxtaposed. The weakening of realism is the abandonment of the ÒuniqueÓ account of a sequence of the events, and its replacement with compatible alternatives, not with a self-consistent collection of all relative properties. 

Once more, this does not mean that $B$ and $A$ cannot communicate their experience. In fact, in either account the possibility of communicating experiences exists and in either account consistency is ensured.  Contradiction emerges only if, against the main stipulation of RQM, we insist on believing that there is an Òabsolute, externalÓ account of the state of affairs in the world, obtained by juxtaposing actualities relative to different observers.

\section{Comparison with Laudisa's discussion of relational EPR}

The EPR argument has been discussed in the context of RQM also by Federico Laudisa, in a recent paper 
\cite{key-11}.  Laudisa's discussion has some points in common 
with the one given here, but it differs from the present one  
 in one key respect. 

Laudisa starts with a reformulation of the EPR hypotheses, 
namely realism, locality and completeness of
QM, in a form meant to be compatible with RQM. 
The locality principle, in particular, is given the following formulation:
\emph{No property of a physical system $S$} \emph{that is objective
relative to some observer can be influenced by measurements performed
in space-like separated regions on a different physical system.} 
He is then able to show that the contradiction between locality,
formulated in this manner, and QM is \emph{itself} relative, in the
sense that it is frame-dependent: there is always an observer (in the
sense of special relativity) for which  it is  inexistent. 

\subsection{Laudisa's Argument}

Here goes Laudisa's argument: after the measurement of the spin of  $\alpha$
(say in the direction $z$) by $A$, the spin of $\beta$ (in the same
direction $z$) has a determined value relative to $A$. However, 
according to the (relativized) locality principle, $\beta$ cannot acquire
a property relative to $A$ as a consequence of the measure performed
on $\alpha$. Hence, relative to $A$, the spin of $\beta$ already had
a determined value \emph{before} the measurement. This fact is in contradiction
with the prior state of the compound system relative to $A$, $\left|\psi\right\rangle_{\alpha+\beta}^{(A)}=\frac{1}{\sqrt{2}}(\left|\downarrow\right\rangle _{\alpha}\left|\uparrow\right\rangle _{\mathbf{\beta}}-\left|\uparrow\right\rangle _{\alpha}\left|\downarrow\right\rangle _{\mathbf{\beta}})$,
which leads to the improper mixture representing the state of $\beta$
relative to $A$: $\rho_{\beta}^{(A)}=\frac{1}{2}(\left|\uparrow\right\rangle _{\beta}\left\langle \uparrow\right|_{\beta}+\left|\downarrow\right\rangle _{\beta}\left\langle \downarrow\right|_{\beta})$.
At that point, Laudisa remarks that, because of space-like separation
between $A$ and $B$, one can find a reference frame in which $A$'s
measurement precedes $B$'s. In such a frame, when $\textrm{A}$ faces
the locality/completeness contradiction, $B$ has not performed any
measurement yet, and therefore escapes the contradiction. What is
more, there exists another reference frame in which the chronology
of measurements is inverted, so that the contradiction afflicts $B$
but not $A$. Finally, the EPR contradiction turns out to be frame-dependent,
and thus fails to refute the locality principle in an absolute sense.

\subsection{Comparison}

Laudisa's interpretation is based on the same premise --relativity
of quantum states--, but differs from the one presented
here. Unlike Laudisa, we do not understand locality
as prohibiting the acquisition of information by an observer on a
distant system, but only as prohibiting the possibility that a
measurement performed in a region 
could, in any way, affect the outcomes of a measurement 
happening in a distant region.  In the EPR
scenario we have discussed, the state of $\beta$ relative to $B$
is independent of $A$'s measurement, but not the state of $\beta$ relative to
$A$. Since the existence of correlations between $\alpha$ and $\beta$
is know \emph{a priori} by $A$, the measurement of an individual
observable of $\alpha$ does permit the prediction of the value of
an individual observable of $\beta$. What is affected by this
measurement is not a hypothetical absolute physical state of $\beta$,
but just $A$'s knowledge about $\beta$. It is  $B$'s
knowledge (or direct experience) about $\beta$ that cannot be affected by anything 
performed by $A$.

Laudisa's residual frame--dependent  contradiction between locality 
and completeness results from an interpretation of locality which 
disregards the epistemic aspect of relative states. 
More radical, our conclusion here is that there is no contradiction 
at all between locality and completeness, nor, more generally, locality
and QM predictions. 

\section{Conclusion}

We have argued that within the relational framework the
EPR-type correlations predicted by QM do not violate
locality.   In fact, the relation between locality and QM 
is more than the ``peaceful coexistence'' which is often declared:
rather, from the relational perspective, QM is rooted in locality in 
a way which, although it dismisses Einstein's strict realism
(the ``real, objective state of affairs''), certainly 
corroborates QM's claim to be a fundamental theory. 

Needless to say, the weakening of realism implied by RQM
may be considered too high a price to pay by some. (This view is strongly
argued, for instance, in the recent \cite{key-99}.)  Our opinion, instead,
is that after almost a century of substantial failure, it may be worthwhile
to try some bold philosophical step, expanding the original motivations of Heisenberg and Bohr, in order to make 
full sense of quantum mechanics. 

Einstein's original motivation with EPR was not to question locality, 
but rather to question the completeness of QM, on the basis of
a firm confidence in locality.  The EPR argument has then been turned 
upside down, and has been perceived as evidence
for non-locality (in fact, a peculiar form of non--locality) in QM,
independently from the issue of completeness: after Bell, indeed, it is generally
assumed that even a hidden variable theory that completes QM \emph{must}
be non-local. RQM is complete in the sense of exhausting 
everything that can be said about nature.  However, in a sense
RQM can be interpreted as the discovery of the incompleteness
of the description of reality that any \emph{single} observer 
can give: $A$ can measure the pointer variable of $B$, but 
the set of the events as described $B$ is irreducibly
distinct from the set of events as described by $A$. In this particular 
sense,
RQM can be said to show the ``incompleteness" of single--observer 
Copenhagen QM. Then  Einstein's intuition that the EPR
correlations reveal something deeply missing in Copenhagen 
quantum mechanics can be understood as being correct:  The  
incompleteness of Copenhagen QM is the disregard of the 
quantum properties of all observers, which leads to paradoxes
as the apparent violation of locality exposed by EPR. 
 
This recalls the conclusion that the late Prof. Peres 
reached in his analysis of EPR in 2004: ``The question
raised by EPR `Can the quantum--mechanical description of physical
reality be considered complete?' has a positive answer. However,
reality may be different for different observers''  \cite{key-34}. This is the
idea at the basis of RQM. 

\vskip.5cm
\centerline{-----------------}
\vskip.5cm
We are grateful to Bas van Fraassen and Alexei Grinbaum for useful comments and suggestions, and to Daniel R.\,Terno for bringing 
reference \cite{key-34} to our attention.
CR would like to dedicate this article to the memory 
of Oreste Piccioni. 

\end{document}